\documentclass[11pt]{article}
\usepackage[utf8]{inputenc}
\usepackage{geometry}
\usepackage{titling}
\usepackage{authblk}
\usepackage{amssymb}
\usepackage{amsmath}
\usepackage{amsthm}
\allowdisplaybreaks[4]
\usepackage{slashed}
\usepackage{graphicx,color,epsfig}
\usepackage{cite}
\usepackage[colorlinks=true,linkcolor=red,citecolor=blue]{hyperref}
\DeclareMathAlphabet{\pazocal}{OMS}{zplm}{m}{n}
\newcommand{\beq}{\begin{eqnarray}}
\newcommand{\eeq}{\end{eqnarray}}

\begin{document}
\title{Charmless Quasi-two-body $B$ Decays in Perturbative QCD Approach:\\
 Taking $B\to K({\cal R}\to) K^+K^-$ As Examples}
\author{Wen-Feng Liu}
\author{Zhi-Tian Zou}
\author{Ying Li$\footnote{liying@ytu.edu.cn}$}
\affil{\it Department of Physics, Yantai University, Yantai 264005, China}
\maketitle
\vspace{0.2cm}

\begin{abstract}
Three-body $B$ decays not only significantly broaden the study of $B$ meson decay mechanisms, but also provide information of resonant particles. Because of complicate dynamics, it is very hard for us to study the whole phase space in a specific approach. In this review, we take $B\to K({\cal R}\to) K^+K^-$ decays \cite{Zou:2020atb} as examples and show the application of the perturbative QCD (PQCD) approach in studying the quasi-two-body $B$ decays, where two particles move collinearly with large energy and the bachelor one recoils back. To describe the dynamics of two collinear particles, the ($S$, $P$ and $D$)-wave functions of kaon-pair with different waves are introduced. By keeping the transverse momenta, all possible diagrams including the hard spectator diagrams and annihilation ones can be calculated in PQCD approach. Most results are well consistent with the current measurements from BaBar, Belle and LHCb experiments. Moreover, under the narrow-width approximation we can extract the branching fractions of the two-body decays involving the resonant states, and also predict the branching fractions of the corresponding quasi-two-body decays $B\to K(\cal{R}\to )\pi^+\pi^-$. All prediction are expected to be tested in the ongoing LHCb and Belle-II experiments.
\end{abstract}
\section{Introduction}
It is well known that $B$ meson hadronic decays provide us unique information on three fronts: $CP$ violation and the angles of the CKM matrix, the structure of QCD in the presence of heavy quarks and energetic light particles, and new particles beyond the standard model (SM) physics in the quark sector. In past twenty years, experimental information on non-leptonic $B$ decays has been collected progressively, at CLEO, two $B$-factories BaBar and Belle \cite{BaBar:2014omp}, the Tevatron, and currently at the LHC, most prominently at LHCb \cite{Aaij:2019hzr}. In addition, the running  Belle II also includes serious plans for analyses of non-leptonic $B$ decays \cite{ Belle-II:2018jsg} with higher precision. Because of large combinatoric backgrounds, studies of charmless $B$ decays have concentrated mainly on two-body decay processes. Although with complex kinematics, three-body decays could not only significantly broaden the study of $B$ meson decay mechanisms and provide additional possibilities for direct $CP$ violation searches, but also provide information of resonant particles. In past decades, more and more analysis of three-body decays have been performed by the BaBar, Belle, CLEO and LHCb, and the branching fractions and $CP$ violations have been measured with high precision. Motivated by the accumulated experimental results, many theoretical studies of various three-body $B$ hadronic decays have been performed in different frameworks, such as approaches based on the symmetry principles \cite{Gronau:2005ax, Engelhard:2005hu, Imbeault:2011jz, Bhattacharya:2013boa, He:2014xha}, the QCD factorization (QCDF) \cite{ElBennich:2009da, Krankl:2015fha, Virto:2016fbw, Cheng:2002qu, Li:2014oca, Li:2014fla, Huber:2020pqb,Zhang:2013oqa, Wang:2015ula, Qi:2018syl, ElBennich:2006yi}, the perturbative QCD approach (PQCD) \cite{Chen:2002th, Wang:2014ira, Wang:2015uea, Li:2016tpn, Wang:2017hao, Rui:2017bgg, Zou:2020atb, Zou:2020ool, Yang:2021zcx}, and other theoretical methods \cite{Liang:2015qva, Ahmed:2020qkv,Shi:2021ste}.

\begin{figure}[!htb]
\begin{center}
\includegraphics[width=7.0cm]{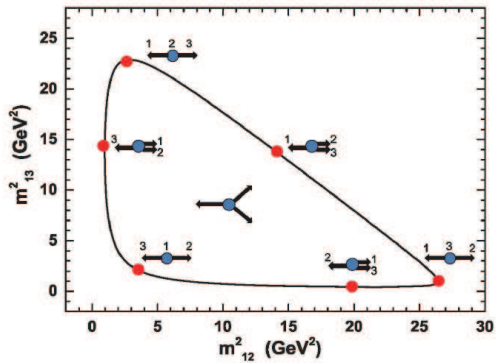}
\caption{Phase space of the three-body decay $B^-\to K^+K^-\pi^-$ in terms of the invariants $m_{12}$ and $m_{13}$ \cite{Cheng:2016shb}.}
\label{DPs}
\end{center}
\end{figure}

Differing from the two-body $B$ decays where the kinematics are fixed, the momentum of each final state in three-body decays is variable. Moreover, both resonant and non-resonant contributions are involved, therefore how to disentangle the resonant and non-resonant contributions reliably is also very important in studying the multi-body decays. For clarity, we define the kinematics of the three-body decay as
\begin{eqnarray}
B (p_B) \to M_a(p_1) M_b(p_2)M_c(p_3).
\end{eqnarray}
It is customary to take these variables as two invariant masses of two pairs of final state particles:
\begin{eqnarray}
m_{12}^2 = (p_1 + p_2)^2\ ,\quad m_{13}^2= (p_1 + p_3)^2.
\end{eqnarray}
Thus, the amplitude of the three-body decay is a function of the two kinematic variables, $m_{12}^2$ and $m_{13}^2$, and the corresponding distribution in the phase-space region is called a Dalitz plot, which can be divided in different regions with characteristic kinematics, as shown in Fig.~\ref{DPs}. As usual, we refer to the $B$-meson rest frame. In the centre of the Dalitz plot, the three final states particles almost have a large energy ($E\sim m_B/3$) and none of them flies collinearly to any others. The corners imply that one final particle is approximately at rest, and the other two fly back-to-back energetically, leading to that one invariant mass is large, and the other two are small. Specially,  the central part of the edges mean that two particles move collinearly with large energy and the other particle recoils back. In the naive factorization, the contribution at the centre of the Dalitz plot is viewed to be both power-suppressed and $\alpha_s$ suppressed with respect to that at the edge, because two hard gluons are need. However, recent studies based on QCD factorization showed that the power corrections are large, which means that the centre region maybe as important as the regions at the edge \cite{Virto:2016fbw, Krankl:2015fha}. Assuming that these different regions can be well described by the respective calculations, it is a very challenging work to determine how well these descriptions merge at intermediate kinematical regimes. So,  the theory of three-body non-leptonic decays is still in an early stage of development.

In the past decades, the $B$ mesons two-body hadronic decays have been explored systematically in the context of QCD factorization, soft-collinear effective theory or PQCD approach. It is natural to ask if the accumulated experience can be used to study the multi-body $B$ decays. Inspired by this, we paid much attention to the edges regions of the Dalitz plot, in which the two energetic particles are collinear and form a moving-fast meson-pair, and the interactions between the meson-pair and the bachelor particle are power suppressed naturally. This kind of process is also called quasi-two-body process. The interactions in the meson-pair can be absorbed into the two-meson wave function. In this case, the obvious generalization of the factorization theorem for two-body decays applies. It is reasonable for us to assume the validity of the factorization for these quasi two-body $B$ decays. In the PQCD framework that is based on the $k_{\rm T}$ factorization, the decay amplitude of quasi-two-body $B$ decays can be decomposed as the convolution \cite{Li:2003yj}
\begin{equation}\label{convention}
\mathcal{A}(B\to M_3(R\to) M_1M_2)=C(t)\otimes\Phi_B\otimes\mathcal{H}\otimes\Phi_{M_1M_2}\otimes\Phi_{M_3}\otimes \exp[-S(t)]
\end{equation}
where the $\Phi_{B}$, $\Phi_{M_3}$ are the wave functions of $B$ meson and the light bachelor meson, respectively. The newly introduced $\Phi_{M_1M_2}$ is the two-meson wave function, which describes how four quarks are combined into two mesons. It should be emphasized that both resonant and nonresonant contributions to the hadron-pair system are included into the wave function. The Wilson coefficient $C(t)$ covers all physics above $m_b$ scale. The exponential term is the so-called Sudakov form factor caused by the additional scale introduced by the intrinsic transverse momenta $k_{\rm T}$, which suppresses the soft dynamics effectively.  The hard kernel $\mathcal{H}$ for the $b$ quark decay, similar to the two-body case, starts with the diagrams of single hard gluon exchange, and it can be calculated perturbatively.

In the following, we will take $B\to K({\cal R}\to) K^+K^-$ decay as examples to show how to study  quasi-two-body processes in the PQCD approach at leading order, where $K^+K^-$-pair move collinearly and combine into resonances with different waves, such as $f_0(980)$, $\phi(1020)$ and $f_2^\prime(1525)$. In Sec.~\ref{frame}, we will introduce the definitions of the two-kaon wave functions with different waves, and present the typical feynmann diagrams in PQCD. In Sec.~\ref{results}, the numerical results and discussions will be given. At last, we summarize this review in Sec.~\ref{summary}.

\section{Framework} \label{frame}
In the quasi-two-body region of phase space, the Dalitz plot analysis allows one to describe the decay amplitude in the isobar model, where the decay amplitude is represented by a coherent sum of amplitudes from  $N$ individual decay channels with different resonances,
\begin{eqnarray} \label{isobar}
\mathcal{A}=\sum_{j=1}^{N} c^j \mathcal{A}^j,
\end{eqnarray}
where the $\mathcal{A}^j$ is the amplitude corresponding to certain resonance and $c^j$ is the complex coefficient describing the relevant magnitude and phase of the different decay channel.

When we calculate the amplitude $\mathcal{A}^j$ in PQCD, the most important inputs are the nonperturbative wave functions. For the decay $B^0\to K^0({\cal R}\to ) K^+K^-$, the wave functions of the $B$ meson and the bachelor meson $K^0$ have been well determined by comparing the theoretical predictions with those well measured experimental data of the two-body $B$ decays. For the sake of brevity, we shall not discuss them anymore, and their explicit expressions can be found in refs.\cite{Keum:2000ph, Lu:2000em, Ali:2007ff}. As aforementioned, in the quasi-two-body decays, the newly introduced ingredient is the two-meson wave function corresponding to different resonances with different spin \cite{Wang:2014ira}.

We firstly discuss the $S$-wave two-meson wave function of the kaon-pair,
\begin{eqnarray}
\Phi_{KK,S}=\frac{1}{\sqrt{2Nc}}[P\mkern-11.5mu/\phi_S(z,\zeta,\omega^2)+\omega\phi_S^s(z,\zeta,\omega^2)
+\omega(n\mkern-9.5mu/ v\mkern-7.5mu/-1)\phi_S^t(z,\zeta,\omega^2)],
\end{eqnarray}
where $z$ is the momentum fraction of the spectator quark, and $\zeta$ is the momentum fraction of one kaon in the pair.$N_c=3$ is the number of colors, and $\omega$ and $P$ are the invariant mass and momentum of the kaon-pair, respectively. $n=(1, 0, {\bf 0_{\rm T}} )$ and $v=(0, 1, {\bf 0_{\rm T}} )$ are two dimensionless vectors. The $\phi_S$, $\phi_S^s$, $\phi_S^t$ are the twist-2 and twist-3 distribution amplitudes, and they are parameterized as \cite{Diehl:1998dk}
\begin{eqnarray}
\phi_S(z,\zeta,\omega^2)&=&\frac{9}{\sqrt{2Nc}}F_S(\omega^2)a_Sz(1-z)(2z-1),\\
\phi_S^s(z,\zeta,\omega^2)&=&\frac{1}{2\sqrt{2Nc}}F_S(\omega^2),\\
\phi_S^t(z,\zeta,\omega^2)&=&\frac{1}{2\sqrt{2Nc}}F_S(\omega^2)(1-2z).
\end{eqnarray}
The Gegenbauer moment $a_S=-0.8$ is determined with the experimental data \cite{Lees:2011nf}. We have to adopt the asymptotic form within the SU(3) symmetry, as the reliable theoretical studies are still absent.

Specially, the function $F_S(\omega^2)$ is the $S$-wave time-like form factor, which contains the interactions between the two kaons in the kaon-pair. For most resonances, the form factors are usually taken to be relativistic Breit-Wigner (RBW) line shapes \cite{Zyla:2020zbs}
\begin{eqnarray}
F_S(\omega^2)=\frac{m_j^2}{m_j^2-\omega^2-im_j\Gamma_j(\omega)},\label{RBW}
\end{eqnarray}
where $m_j$ is the mass of the resonance, and $\Gamma(\omega)$ is the mass-dependent width \cite{Blatt:1952ije, Zyla:2020zbs}. However, the case of $f_0(980)$ is complicated. Because there is an anomalous structure corresponding to the enhancement from the $KK$ system found around $980$ MeV in the $\pi^+\pi^-$ scattering \cite{AlstonGarnjost:1971kv, Flatte:1972rz}, it can be interpreted as a two-channel resonance combining the $\pi\pi$ and $KK$ channels. In order to include these effects, eq.(\ref{RBW}) is modified to the Flatt\'{e} form \cite{Flatte:1976xu, Bugg:2008ig, Aaij:2014emv} as
\begin{eqnarray}\label{Flatte form}
F_S(\omega^2)=\frac{m_{f_0(980)}^2}{m_{f_0(980)}^2-\omega^2
-im_{f_0(980)}(g_{\pi\pi}\rho_{\pi\pi}+g_{KK}\rho_{KK}F_{KK}^2)},
\end{eqnarray}
where $g_{\pi\pi}$ and $g_{KK}$ are the $f_0(980)$ coupling constants to the $\pi\pi$ and $KK$ final states, respectively. The phase space factors $\rho_{\pi\pi}$ and $\rho_{KK}$ are given as
\begin{eqnarray}
\rho_{\pi\pi}=\sqrt{1-\frac{4m_\pi^2}{\omega^2}},\,\,\,\,
\rho_{KK}=\sqrt{1-\frac{4m_K^2}{\omega^2}}.
\end{eqnarray}
The factor $F_{KK}=e^{-\alpha q^2}$ with $\alpha \approx 2.0 ~\rm GeV^{-2} $ is introduced to suppress the $K\overline K$ contribution \cite{Aaij:2014emv}, $q$ being the momentum of each kaon in the  $K\overline K$ rest frame.

It should be noted that we only consider the contributions from three scalar resonances $f_0(980)$, $f_0(1500)$ and $f_0(1710)$, which have been well analyzed by BaBar experiments \cite{Lees:2012kxa,Lees:2011nf}. The coefficients of the coherence summation of these three resonances in eq.(\ref{isobar}) are set to be $c^{f_0(980)}=2.9$, $c^{f_0(1500)}=1.0$, $c^{f_0(1710)}=0.5$, which have been determined by the experimental measurements already \cite{Lees:2012kxa, Lees:2011nf}.

Next, we come to the $P$-wave two-kaon wave function. Because the third bachelor kaon in $B^0\to K^0 K^+K^-$ decay is a pseudoscalar one, therefore only the longitudinal part contributes, and its form can be expressed as
\begin{eqnarray}
\Phi_{KK,P}=\frac{1}{\sqrt{2N_c}}\left(\not \! P\phi_P(z,\zeta,\omega)+\omega\phi_P^s(z,\zeta,\omega)
+\frac{\not \!p_1\not \!p_2-\not \!p_2\not \!p_1}{\omega(2\zeta-1)}\phi_P^t(z,\zeta,\omega)\right),
\end{eqnarray}
where $p_{1(2)}$ is the momentum of each kaon in the $K\bar K$-pair. The corresponding twist-2 and 3 distribution amplitudes that can be decomposed as Gegenbauer polynomials are given as
\begin{eqnarray}
\phi_P^0(z,\zeta,\omega)&=&\frac{3F_P^{\parallel}(\omega^2)}{\sqrt{2N_c}}z(1-z)\Big[1+a_P^0C_2^{3/2}(2z-1)\Big](2\zeta-1),\\
\phi_P^s(z,\zeta,\omega)&=&\frac{3F_P^{\perp}(\omega^2)}{2\sqrt{2N_c}}(1-2z)\Big[1+a_P^s(1-10z+10z^2)\Big](2\zeta-1),\\
\phi_P^t(z,\zeta,\omega)&=&\frac{3F_P^{\perp}(\omega^2)}{2\sqrt{2N_c}}(2z-1)^2\Big[1+a_P^tC_2^{3/2}(2z-1)\Big](2\zeta-1),
\end{eqnarray}
with $a_P^0=-0.6$, $a_P^s=-0.8$, and $a_P^t=-0.3$. Similarly, the $P$-wave time-like form factor $F_P^{\parallel}(\omega^2)$ is also taken to be the RBW line shape in eq. (\ref{RBW}). As for  $F_P^{\perp}(\omega^2)$, the relation \cite{Wang:2016rlo}
 \begin{eqnarray}
  \frac{F_P^{\parallel}(\omega^2)}{F_P^{\perp}(\omega^2)}\approx \frac{f_V}{f_V^T},
 \label{perp}
 \end{eqnarray}
will be used, where $f_V$ and $f_V^T$ are the vector and tensor decay constants.

At last, we will discuss the wave function of $D$-wave kaon-pair in which the information of tensor meson resonances is included. In previous studies of $B$ meson decays involving a tensor \cite{Zou:2012td,Cheng:2010yd},it is found that the polarization components $\pm2$ of tensor meson do not contribute of the amplitudes, due to the conservation of the angular momentum. Hence, the form of $D$-wave two-kaon wave function is the same as that of the $P$-wave,
 \begin{eqnarray}
 \Phi_{KK,D}=\frac{1}{\sqrt{2N_c}}\left(p\mkern-8.5mu/\phi_D(z,\zeta,\omega)+\omega\phi_D^s(z,\zeta,\omega)
+\frac{p\mkern-8.5mu/_1p\mkern-8.5mu/_2-p\mkern-8.5mu/_2p\mkern-8.5mu/_1}{\omega(2\zeta-1)}\phi_D^t(z,\zeta,\omega)\right).
\end{eqnarray}
The distribution amplitudes are given as
\begin{eqnarray}
\phi_D(z,\zeta,\omega)
&=&\sqrt{\frac{2}{3}}\frac{9F_D^{\parallel}(\omega^2)}{\sqrt{2N_c}}z(1-z)a_D^0\Big[2z-1\Big]P_2(\zeta),\\
\phi_D^s(z,\zeta,\omega)
&=&-\sqrt{\frac{2}{3}}\frac{9F_D^{\perp}(\omega^2)}{4\sqrt{2N_c}}a_D^0\Big[1-6z+6z^2\Big]P_2(\zeta),\\
\phi_D^t(z,\zeta,\omega)
&=&\sqrt{\frac{2}{3}}\frac{9F_D^{\perp}(\omega^2)}{4\sqrt{2N_c}}a_D^0(2z-1)\Big[1-6z+6z^2\Big]P_2(\zeta),
\end{eqnarray}
with $a_D^0=0.6$ and $P_2(\zeta)=1-6\zeta+6\zeta^2$. Likewise, we use the RBW line shape for describing the time-like form factor $F_D^{\parallel}$ and determine the $F_D^{\perp}$ by the similar relation as eq.(\ref{perp}).

%===========================================================================
\begin{figure}[!htb]
\begin{center}
\includegraphics[scale=0.5]{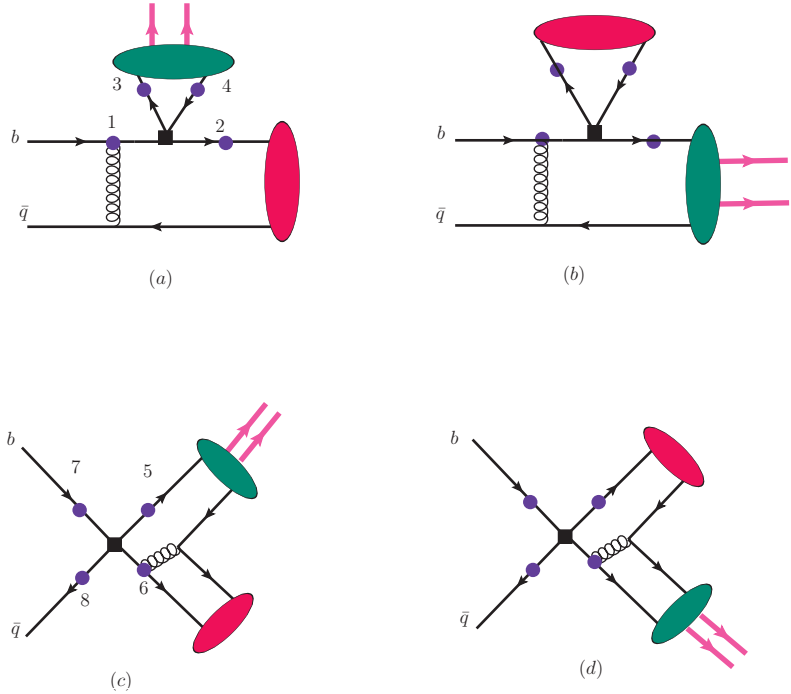}
\caption{Typical Feynman diagrams for the quasi-two-body decay $B \to K ({\cal R}\to) K^+ K^-$ in PQCD, where the black squares stand for the weak vertices, large (purple) spots on the quark lines denote possible attachments of hard gluons, and the green ellipse represent Kaon-pair and the red one is the light bachelor $K$ meson\cite{Zou:2020atb}.}\label{feynman}
\end{center}
\end{figure}
%============================================================================

According to the effective Hamiltonian of $b$ quark decay, we can draw the possible Feynman diagrams as shown in Fig.~\ref{feynman}. In diagram (a), the spectator quark enters to the bachelor particle, and the $KK$-pair is emitted. In order to make the spectator quark from soft to hard, the hard gluon is needed, and it can come from any quark participating in the four-quark weak interaction. So, with three wave functions, we could calculate amplitude with different operators in the PQCD approach. In practice, we keep the transverse momenta of each quark for smearing the end-point singularities. For decays $B^\pm\to  K^\pm({\cal R}\to) K^+ K^-$, the spectator quark can also enter into the recoiled meson pair, as shown in diagram (b). Besides these two kinds of contributions, the annihilation diagrams (c) and (d) also play important roles, which can be calculated quantitatively without introducing new parameters. Due to the space limit, we cannot list all amplitudes of these diagrams, and one can find them in ref.\cite{Zou:2020atb}.

\section{Results and Discussions} \label{results}
With the calculated amplitude $\mathcal{A}$, we then obtain the differential branching fraction
\begin{eqnarray}
\frac{d^2\mathcal{B}}{d \zeta d\omega}=\frac{\tau_B\omega|\vec{p}_1||\vec{p}_3|}{32\pi^3 m_{B}^3}|\mathcal{A}|^2,
\end{eqnarray}
where $\tau_B$ is the life-time of the $B$ meson. The magnitudes of three-momenta of one kaon and the bachelor particle in the rest frame of the $KK$-pair are given by in the rest frame of the $KK$-pair are given by
\begin{eqnarray}
|\vec{p}_1|=\frac{\sqrt{\lambda(\omega^2,m_K^2,m_K^2)}}{2\omega}, \quad
|\vec{p}_3|=\frac{\sqrt{\lambda(m_{B}^2,m_K^2,\omega^2)}}{2\omega},
\end{eqnarray}
with the standard K$\ddot{a}$ll$\acute{e}$n function $\lambda (a,b,c)= a^2+b^2+c^2-2(ab+ac+bc)$.

Integrating out the variables $\zeta$ and $\omega$, we could obtain the branching fractions of the quasi-two-body decays $B\to K({\cal R} \to) K^+K^-$, and the $CP$ averaged branching fractions are presented in Table.\ref{br}. As aforementioned, the theoretical studies of multi-body $B$ hadronic decays are on the early stage, so there are many uncertainties in our calculations. The first and foremost parameters uncertainties are from nonperturbative parameters, including the shape parameter in the heavy $B$ wave function, the Gegenbauer moments in the distribution amplitudes of kaon-pair and kaon. We call for the results with higher precision from nonperturbative approaches, such as QCD sum rules and lattice QCD.  The second uncertainties origin from the PQCD approach itself, including the unknown QCD radiative corrections and the power corrections. Although some higher order and higher power corrections have been explored, the complete calculations are still in absence. The last kind of uncertainties are caused by the CKM matrix elements. In the table, the currently available experimental measurements are also given for comparison. It is obvious that within the uncertainties most of our results are in good agreement with experimental results, except decay mode $B^0 \to K^0(f_0(1710)\to ) K^+ K^-$

\begin{table}[htb!]
\caption{$CP$ averaged branching fractions (in $10^{-6}$) of $B\to K({\cal R} \to) K^+K^-$ decays in PQCD \cite{Zou:2020atb} together with experimental data~\cite{Zyla:2020zbs}. } \label{br}
\begin{center}
\begin{tabular}{l c c }
 \hline \hline
 \multicolumn{1}{c}{Decay Modes}&\multicolumn{1}{c}{PQCD } &\multicolumn{1}{c}{EXP\cite{Zyla:2020zbs}}   \\
\hline\hline
 $B^+ \to K^+(\phi(1020)\to )K^+ K^-$
 &$3.81^{+1.44+0.64+0.27}_{-1.03-0.33-0.00}$
 &$4.48\pm0.22^{+0.33}_{-0.24}$\\

 $B^+ \to K^+(f_0(980)\to) K^+ K^-$
 &$10.13^{+5.60+2.22+0.71}_{-4.38-2.44-0.00}$
 &$9.4\pm1.6\pm2.8$\\

 $B^+ \to K^+(f_0(1500)\to) K^+ K^-$
 &$0.60^{+0.24+0.07+0.05}_{-0.24-0.06-0.02}$
 &$0.74\pm0.18\pm0.52$\\

 $B^+ \to K^+(f_0(1710)\to) K^+ K^-$
 &$1.64^{+0.89+0.42+0.08}_{-0.70-0.46-0.02}$
 &$1.12\pm0.25\pm0.50$\\

 $B^+ \to K^+(f_2^{\prime}(1525)\to) K^+ K^-$
 &$0.68^{+0.37+0.13+0.07}_{-0.29-0.14-0.00}$
 &$0.69\pm0.16\pm0.13$\\

 $B^+ \to K^+(f_2(2010)\to) K^+ K^-$
 &$1.18^{+0.65+0.26+0.12}_{-0.50-0.19-0.00}$
 &$$
\\
 \hline
  $B^0 \to K^0(\phi(1020)\to)K^+ K^-$
  &$3.22^{+1.36+0.48+0.18}_{-0.98-0.18-0.08}$
  &$3.48\pm0.28^{+0.21}_{-0.14}$\\

  $B^0 \to K^0(f_0(980)\to) K^+ K^-$
  &$9.10^{+5.12+2.19+0.69}_{-3.89-2.11-0.00}$
  &$7.0^{+2.6}_{-1.8}\pm2.4$\\

  $B^0 \to K^0(f_0(1500)\to) K^+ K^-$
  &$0.57^{+0.26+0.09+0.04}_{-0.22-0.15-0.00}$
  &$0.57^{+0.25}_{-0.19}\pm0.12$\\

  $B^0 \to K^0(f_0(1710)\to) K^+ K^-$
  &$1.48^{+0.82+0.39+0.11}_{-0.63-0.42-0.00}$
  &$4.4\pm0.7\pm0.5$\\

  $B^0 \to K^0(f_2^{\prime}(1525)\to) K^+ K^-$
  &$0.58^{+0.31+0.12+0.05}_{-0.27-0.13-0.01}$
  &$0.13^{+0.12}_{-0.08}\pm0.16$
  \\

  $B^0 \to K^0(f_2(2010)\to) K^+ K^-$
  &$1.09^{+0.57+0.26+0.11}_{-0.48-0.23-0.00}$
  &
   \\

 \hline \hline
\end{tabular}
\end{center}
\end{table}

\begin{figure}[htb!]
\begin{center}
\includegraphics[scale=0.5]{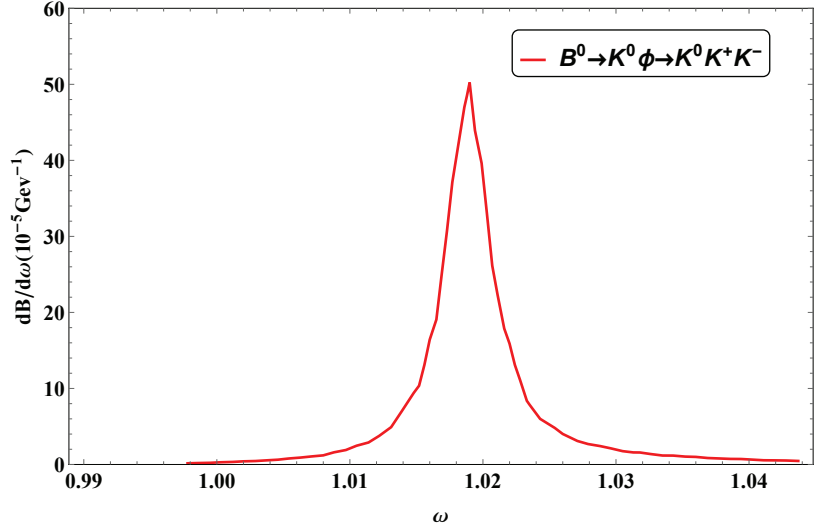}
\caption{The $\omega$-dependence of differential branching fractions for the $B^0\to K^0K^+K^-$ decay \cite{Zou:2020atb}.}\label{Fig:2}
\end{center}
\end{figure}

In Fig.~\ref{Fig:2}, we show the $K^+K^-$ invariant mass-dependent differential branching fractions for the quasi-two-body decays $B^0\to K^0(\phi(1020) \to ) K^+K^-$. It can be seen that the contribution of $\phi(1020)$ is dominate in the region near 1 GeV. Under the narrow-width approximation, the three-body decay and corresponding two-body one satisfy the relation
\begin{eqnarray}
\mathcal{B}(B\to P_3 ({\cal R}\to)  P_1P_2)\approx\mathcal{B}(B\to P_3 {\cal R})\times\mathcal{B}({\cal R}\to P_1P_2),
\label{nwa}
\end{eqnarray}
with ${\cal R}$ being the resonance. With the experimental results $\mathcal{B}(\phi\to K^+K^-)=(49.2\pm0.5)\%$ \cite{Zyla:2020zbs}, we obtain ${\cal B}(B^0\to K^0\phi)=(6.4^{+2.9}_{-2.0})\times 10^{-6}$, which is in agreement with above experimental results with uncertainties.

\begin{figure}[htb!]
\begin{center}
\includegraphics[scale=0.5]{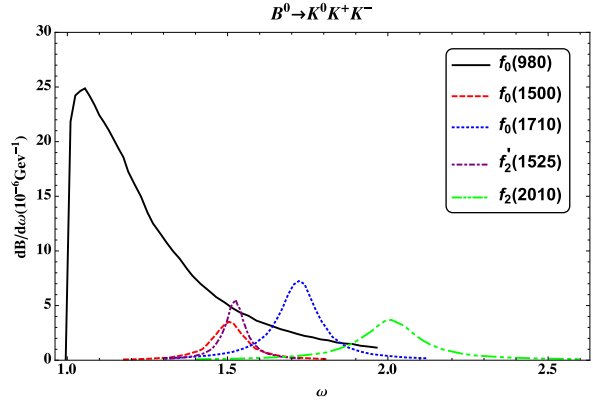}
\caption{The $\omega$-dependence of differential branching fractions from $f_0(980)$, $f_0(1500)$, $f_0(1710)$, $f_2^\prime(1525)$ and $f_2(2010)$ for the $B^0\to K^0K^+K^-$ decay \cite{Zou:2020atb}.}\label{Fig:3}
\end{center}
\end{figure}

Although the conventional quark model has achieved great success, the quark structures of scalar particles are still quite controversial, especially for the light scalar ones. Although there are many hints that the light scalars are four-quark states, the wave functions of these particle have not established, and we left it as our future work. Currently, we here still regard $f_0(980)$ as a two quark structure. Many experimental evidences indicate that both $s\bar s$ and $q\bar q$ are involved in the $f_0(980)$, so the mixing form is supposed to be
\begin{eqnarray}
|f_0(980)\rangle=|q\bar{q}\rangle\sin\theta+|s\bar{s}\rangle\cos\theta,
\end{eqnarray}
with $q\bar{q}=(u\bar{u}+d\bar{d})/\sqrt{2}$ and the mixing angle $\theta=40^{\circ}$ \cite{Cheng:2002ai}. For the heavier scalar mesons, there are glueball contents in isosinglet scalar mesons $f_0(1370)$, $f_0(1500)$ and $f_0(1710)$. It is commonly accepted that $f_0(1710)$ is dominated by the scalar glueball, while $f_0(1500)$ is an approximately SU(3) octet with negligible glueball component. So, the glueball content of $f_0(1500)$ can be neglected.

In Fig.~\ref{Fig:3}, we show the dependencies of the differential branching ratios $d{\cal B}(B^+\to K^+({\cal R}\to) K^+K^-)/d\omega$ on the kaon-pair invariant mass $\omega$  with the $S$-wave resonances $f_0(980)$, $f_0(1500)$ and $f_0(1710)$ and the $D$-wave particles $f_2^\prime(1525)$ and $f_2(2010)$. From the figure, it is obvious that the $f_0(980)$ production is apparently dominant, and it is about ten times larger than that of $f_0(1710)$. In fact, it has been shown that within present experimental and theoretical uncertainties, the narrow peak at 980 MeV in the $K\bar K$ system may be interpreted as a cusp phenomenon with large width, and does not necessarily imply that there exists a narrow resonance in that region. Because of large width, the contribution of the tail of $f_0(980)$ with is still larger than the total effect of $f_0(1500)$. We also find that the contributions of $f_0(1710)$ and  $f_0(1500)$ overlap with each other, so that it is very hard for us to disentangle two resonances at the region about $1.5~\rm GeV$. Unlike $S$-wave, the contributions from $D$-wave resonances do not overlap anymore because of the narrow width of $f_2^\prime(1525)$. As for the $f_2 (1270)$, although there are some studies on it \cite{Li:2018lbd}, the mixing between different quark components is quit complicated, so its contribution has not been included.

As aforementioned, the narrow-width approximation is invalid in describing the scalar particle $f_0(980)$, but it  still holds in the processes $B \to K(f_0(1500) \to) K^+K^-$. Again, we could obtain the branching fractions of $B\to Kf_0(1500)$ as
 \begin{eqnarray}
 \mathcal{B}(B^{0}\to K^{0}f_0(1500))&=&(13.7\pm6.1)\times 10^{-6},\\
 \mathcal{B}(B^{+}\to K^{+}f_0(1500))&=&(13.9\pm5.8)\times 10^{-6},
 \end{eqnarray}
within ${\cal B}(f_0(1500)\to K^+K^-)\simeq 4.3\%$. For the decay $B^{0}\to K^{0}f_0(1500)$, our result agrees with experimental data $(1.3^{+0.7}_{-0.6})\times 10^{-5}$ \cite{Zyla:2020zbs}.  As for the decay $B^{+}\to K^{+}f_0(1500)$, the center value of prediction is about 3.7 times larger than that averaged experimental data $ (3.7 \pm 2.2)\times 10^{-6}$ \cite{Zyla:2020zbs}, but both theoretical and experimental uncertainties are very larger.

Under the narrow-width approximation, we get a  ratio as
\begin{eqnarray}
{R}_1=\frac{\mathcal{B}(f_0(1500)\to K^+K^-)}{\mathcal{B}(f_0(1500)\to \pi^+\pi^-)}\approx\frac{\mathcal{B}(B\to K(f_0(1500)\to) K^+K^-)}{\mathcal{B}(B \to K (f_0(1500)\to  \pi^+\pi^-))}.
\end{eqnarray}
Using the experimental data $\mathcal{B}(f_0(1500)\to K^+K^-)=4.3\%$ and $\mathcal{B}(f_0(1500)\to \pi^+\pi^-)=23.27\%$ \cite{Zyla:2020zbs}, we can get the fraction ${R}_1=0.185$. Thereby, the branching fractions of $B\to K(f_0(1500) \to  \pi^+ \pi^-)$ decays are estimated to be
 \begin{eqnarray}
 \mathcal{B}(B^{+}\to K^{+}(f_0(1500)\to)\pi^+\pi^-)&=&(3.24\pm 1.35)\times10^{-6},\\
 \mathcal{B}(B^{0}\to K^{0}(f_0(1500)\to) \pi^+\pi^-)&=&(3.15\pm1.40)\times10^{-6}.
\end{eqnarray}
Similarly, we also define another ratio as
\begin{eqnarray}
{R}_2=\frac{\mathcal{B}(f_0(1710)\to K^+K^-)}{\mathcal{B}(f_0(1710)\to \pi^+\pi^-)}\approx
\frac{\mathcal{B}(B\to K(f_0(1710)\to) K^+K^-)}{\mathcal{B}(B\to K(f_0(1710)\to) \pi^+\pi^-)}.
\end{eqnarray}
With the experimental result $\Gamma (f_0(1710) \to \pi \pi)/\Gamma(f_0(1710)\to K\overline{K})=0.23\pm0.05$ \cite{Zyla:2020zbs}, we then  predict the branching fractions of $B \to K (f_0(1710)\to )\pi^+\pi^-$ decays as
\begin{eqnarray}
 \mathcal{B}(B^{+}\to K^{+}(f_0(1710)\to)\pi^+\pi^-)
&=&(5.0^{+3.9}_{-3.4})\times10^{-7},\nonumber\\
 \mathcal{B}(B^{0}\to K^{0}(f_0(1710)\to)\pi^+\pi^-)
&=&(4.5^{+3.9}_{-3.4})\times10^{-7}.
\end{eqnarray}
All above  predictions are expected to be measured in LHCb and Belle-II experiments.
\section {Summary}\label{summary}
In this review, we took $B\to K({\cal R}\to) K^+K^-$ decays as examples and showed the application of PQCD in studying the quasi-two-body $B$ decays. In this approach, we here only discussed the region at the central part of the edges, where two particles move collinearly with large energy and the other particle recoils back. In order to describe the dynamics of two collinear particles, the wave functions of kaon-pair with different spins are introduced. By keeping the transverse momenta, all possible diagrams at leading order were calculated, including the hard spectator diagrams and annihilation ones. Most of our numerical results are well consistent with the current experimental data. Under the narrow-width approximation we extracted the branching fractions of the corresponding two-body decays involving the resonances, such as the $B \to K \phi $, whose branching fractions agree with the current experimental data well. We also predicted the branching fractions of the corresponding quasi-two-body decays $B\to K(\cal{R}\to) \pi^+\pi^-$. All prediction are expected to be tested in the ongoing LHCb and Belle-II experiments.

%============================================================================
\section*{Acknowledgement}
Y.Li thanks Q-X Li and X Liu for collaborations. This work was supported in part by the National Natural Science Foundation of China under the Grant No. 11975195, by the Natural Science Foundation of Shandong province under the Grant No.ZR2019JQ04, and  by the Project of Shandong Province Higher Educational Science and Technology Program under Grants No. 2019KJJ007.

{\small
\bibliographystyle{bibstyle}
\bibliography{mybibfile}}
\end{document}